\documentclass[conference]{IEEEtran}

\IEEEoverridecommandlockouts

\usepackage{graphics} 
\usepackage{tabularx} 
\usepackage{epsfig} 
\usepackage{mathptmx} 
\usepackage{amsmath} 
\usepackage{amssymb}  
\usepackage{float}
\usepackage{algorithm2e}
\usepackage{multirow}
\usepackage{enumitem}
\usepackage{xcolor} 
\usepackage{cite} 
\usepackage{dblfloatfix} 
\usepackage{url} 

\usepackage{tikz}

\newif\ifarxiv
\arxivtrue             

\newcommand{\copyrighttext}{%
  \footnotesize \textcopyright~2025 IEEE. This paper has been accepted by the 2025 International Conference on Cyber-physical Social Intelligence (CPSI 2025). Personal use of this material is permitted. Permission from IEEE must be obtained for all other uses, in any current or future media, including reprinting/republishing this material for advertising or promotional purposes, creating new collective works, for resale or redistribution to servers or lists, or reuse of any copyrighted component of this work in other works.%
}

\newcommand{\copyrightnotice}{%
  \begingroup
  \newsavebox{\cpybox}
  \sbox{\cpybox}{%
    \fbox{\parbox{\dimexpr\textwidth-\fboxsep-\fboxrule\relax}{\copyrighttext}}%
  }%
  \IEEEpubid{%
    \parbox{\columnwidth}{\vspace{\dimexpr\ht\cpybox+\dp\cpybox+10pt\relax}}%
    \hspace{\columnsep}\makebox[\columnwidth]{}%
  }%
  \begin{tikzpicture}[remember picture,overlay]
    \node[anchor=south, yshift=10pt] at (current page.south)
      {\makebox[\textwidth][c]{\usebox{\cpybox}}};
  \end{tikzpicture}%
  \endgroup
}

\def\BibTeX{{\rm B\kern-.05em{\sc i\kern-.025em b}\kern-.08em
    T\kern-.1667em\lower.7ex\hbox{E}\kern-.125emX}}
\begin{document}

\title{Urban Air Mobility: A Review of Recent Advances in Communication, Management, and Sustainability}

\author{\IEEEauthorblockN{Zhitong He}
\IEEEauthorblockA{\textit{Elmore Family School of ECE} \\
\textit{Purdue University}\\
Indianapolis, USA \\
he733@purdue.edu}
\and
\IEEEauthorblockN{Zijing Wang}
\IEEEauthorblockA{\textit{Global Sustainability Center} \\
\textit{American Bureau of Shipping}\\
Spring, USA \\
zijwang@eagle.org}
\and
\IEEEauthorblockN{Lingxi Li$^*$}
\IEEEauthorblockA{\textit{Elmore Family School of ECE} \\
\textit{Purdue University}\\
Indianapolis, USA \\
lingxili@purdue.edu}

\thanks{The views, findings, and conclusions expressed in this paper are solely those of the authors and do not necessarily represent the official positions or policies of American Bureau of Shipping (ABS).}
}

\maketitle
\copyrightnotice

\begin{abstract}

Urban Air Mobility (UAM) offers a transformative approach to addressing urban congestion, improving accessibility, and advancing environmental sustainability. Rapid progress has emerged in three tightly linked domains since 2020: (1) Communication, where dynamic spectrum allocation and low-altitude channel characterization support reliable air-ground data exchange; (2) UAM management, with novel air-traffic control concepts for dense, largely autonomous urban airspace; and (3) Sustainability, driven by energy-efficient propulsion, integrated charging infrastructure, and holistic environmental assessment. This paper reviews and synthesizes the latest research across these areas, compares the state-of-the-art solutions, and outlines the technological and infrastructural milestones that are critical to realizing a scalable, sustainable UAM ecosystem.
\end{abstract}

\section{Introduction}
Due to the expansion of traffic volume, the amount of travel delays experienced by commuters reached 54 hours per year and is expected to continuously increase \cite{lasley2023report}\cite{usdot2024}. Meanwhile, the growth and development of electric vertical take-off and landing (eVTOL) aircraft technology and the corresponding operational low-altitude airspace aroused the attention of traffic regulators\cite{faa2023evtol}. Urban Air Mobility (UAM), which utilizes low-altitude airspace, has emerged as a revolutionary concept aimed at transforming urban transportation systems \cite{huang2024potential}. By leveraging advanced aerial vehicles, UAM proposes to offer efficient, sustainable, and rapid transportation solutions, thereby addressing the growing concerns related to urban congestion, pollution, and transit efficiency. The UAM market has a huge and promising potential based on the market analysis \cite{cohen2021urban}. The evolution of UAM is closely tied to significant advancements in aerospace technology, electric propulsion, and automation, enabling the development of vehicles capable of vertical takeoff and landing (VTOL). The improvements in battery technology and innovations in drone and autonomous flight technologies have all played pivotal roles in making UAM a feasible proposition \cite{cohen2021urban}.

From an industrial perspective, the focus has been on designing and testing VTOL aircraft, with significant efforts directed toward electric and hybrid propulsion systems to ensure environmental sustainability. Companies like Airbus, Boeing, and new entity such as Joby Aviation have been at the forefront, developing prototypes and conducting test flights \cite{airbus}\cite{boeing}. From the academic perspective, the key areas of research interests include vehicle design and efficiency, air traffic management systems, safety and certification processes, and the societal impacts of integrating UAM into existing urban landscapes \cite{straubinger2020overview}.

Despite recent progress, the path to the realization of UAM on a scale still faces significant challenges. Technical hurdles such as battery life, noise pollution, and vehicle safety remain significant concerns. In addition, the regulatory frameworks for UAM operations are still in the developmental stages, which require extensive collaboration between industry stakeholders and government agencies to ensure safe and efficient airspace integration \cite{cohen2021urban}.


As UAM continues to evolve, it has the potential to radically transform urban landscapes by offering a new dimension of mobility. The ongoing efforts in the industrial and academic sectors are setting the foundation for a future where UAM enhances the livability of cities while addressing critical transportation challenges. However, recent literature reviews focus either on airspace design \cite{yang2024airspace} \cite{bauranov2021designing} or autonomous aerial mobility \cite{mishra2023autonomous} in UAM. Few works have been conducted to emphasize and conclude the ongoing development of UAM technologies. This paper presents recent technological developments that foster the future of UAM. The major contributions of this paper are summarized as follows:
\begin{enumerate}
    \item An in-depth examination of the current technological advancements in UAM was conducted, focusing on the key innovations and developments in communication systems and air traffic management, along with the impacts on transportation sustainability.
    \item The classification of the diverse technological research efforts in the UAM was presented through a comprehensive analysis of the existing literature.
    \item A forward-looking perspective on the challenges and opportunities for UAM was provided.
\end{enumerate}

The remainder of this paper is organized as follows: A literature review of recent research on UAM communication is introduced in Section~\ref{Sec:communication}. Section~\ref{Sec:management} describes the UAM management techniques. Section~\ref{Sec:sustainability} shows how UAM facilitates the growth of sustainability in the transportation system. Section~\ref{Sec:discussion} presents a follow-up discussion based on the research findings. Finally,  Section~\ref{Sec:conclusion} draws the conclusion.

\section{Communication Technologies in UAM}
\label{Sec:communication}

With increasing traffic data demands and integration challenges, a robust and reliable communication system would be essential in UAM operations. Namuduri et al. \cite{namuduri2022advanced} pointed out research directions for Advanced Air Mobility (AAM) and UAM focusing on the fields of communications, navigation, and surveillance, which include topics such as Air Corridors, Air-to-Air Communications, 3GPP (3rd Generation Partnership Project) and Support for Navigation. The researchers emphasized the role of communication in enabling safe navigation, collision avoidance, and coordination with ground systems. Novel vehicular communication techniques and protocols such as cellular vehicle-to-everything (C-V2X) communication \cite{he2024rampcast} can also be applied to aerial communication and UAM operations. In addition, emerging technologies such as 6G communication would bring challenges and opportunities to enhance communication performance in UAM. In this section, we will focus on the current communication infrastructures and the advanced technologies that are under development. Table~\ref{tab:UAM_Communication} presents a brief overview of communication technology, infrastructure levels, and performance metrics for UAM/AAM.

\begin{table*}[htbp]
    \centering
    \caption{Summary of UAM Communication Technologies}
    \resizebox{\textwidth}{!}{
    \begin{tabular}{|c|c|c|c|}
        \hline
        \textbf{Reference} & \textbf{Technology Categories} & \textbf{Infrastructure Levels} & \textbf{Performance Metrics} \\
        \hline
        Zeng \cite{zeng2023wireless} & AI-based Learning & Ground & Connectivity Probability, Model Convergence \\
        \hline
        Zaid \cite{zaid2023evtol} & Communication Infrastructure & Ground/Airborne & Network Coverage, Latency \\
        \hline
        Han \cite{han2023joint} & Spectrum Management & Airborne & Mission Completion Time, Spectrum Allocation Efficiency \\
        \hline
        Hu \cite{hu2022communications} & Experimental Field Testing & Airborne & Signal Loss, Frame Outage Probability \\
        \hline
        Schurwanz \cite{schurwanz2023compressed} & Radar-based Sensing & Ground/Airborne & Imaging Performance, Signal Dynamics \\
        \hline
        Jeong \cite{jeong2024geometric} & Radar-based Navigation & Ground/Airborne & Range Accuracy, Doppler Estimation \\
        \hline
        Pham \cite{pham2023High} & Antenna Design & Airborne & Gain, Azimuth Coverage, Radiation Efficiency \\
        \hline
        Han \cite{han2022deep} & Spectrum Management & Airborne & Spectrum Utilization, Interference Reduction \\
        \hline
        Al-Rebaye \cite{al-rubaye2024} & 6G Communication & Ground/Airborne/Satellite & Latency, Bandwidth, Connectivity \\
        \hline
    \end{tabular}
    }
    \label{tab:UAM_Communication}
\end{table*}

\vspace{-1mm}
\subsection{Communication Infrastructures and Technologies}
\subsubsection{Existing and Emerging Infrastructures}

UAM depends on a robust communication infrastructure to support seamless operations and integration with existing air traffic systems. To address the growing demands for eVTOL vehicles, Zaid et al. \cite{zaid2023evtol} emphasized the importance of a hybrid communication system combining terrestrial and aerial base stations (BSs). This approach leverages the strengths of both types of BSs to ensure reliable coverage, reduced latency, and improved network performance in densely populated urban areas.

Future communication frameworks must accommodate increasing digitization and autonomy. Al-Rebaye et al. \cite{al-rubaye2024} proposed a unified architecture integrating ground-, air-, and satellite-based infrastructures to enable sustainable UAM operations. This architecture positions 6G as a pivotal technology for handling high data traffic demands and ensuring seamless connectivity between UAM vehicles, air traffic control, and other systems. A link budget analysis conducted for green urban environments highlighted both challenges and opportunities for deploying 6G, underscoring its potential to transform UAM communication.

Furthermore, Zeng et al. \cite{zeng2023wireless} developed a spatial model to evaluate wireless connectivity in UAM systems, focusing on the signal-to-interference ratio (SIR)-based connectivity probability between ground base stations and VTOL aircrafts. By integrating this analysis with an asynchronous federated learning (AFL) framework, the study demonstrated improvements in turbulence prediction and operational safety, paving the way for more intelligent communication systems in UAM.

\subsubsection{Spectrum Management}
Efficient spectrum management is critical to ensure reliable communication in UAM systems, particularly given the challenges of spectrum scarcity and potential interference with terrestrial networks. Han et al. \cite{han2022deep} introduced the concept of cellular UAM (cUAM), which integrates UAM operations into existing cellular networks to provide reliable air-to-ground communication. This approach requires advanced dynamic spectrum management strategies to minimize interference and optimize resource allocation. Furthermore, the same research group proposed a joint optimization framework for velocity and spectrum management in UAM systems \cite{han2023joint}. Using a deep reinforcement learning (DRL)-based VD3QN algorithm, the framework enables cooperative learning among aerial vehicles for dynamic spectrum allocation, minimizing mission completion time while improving overall system efficiency. The simulation results validated the algorithm's superior performance compared to the heuristic and baseline methods, showcasing its potential in high-density UAM operations.

\subsubsection{Hardware Considerations}

Effective hardware design is fundamental to ensuring reliable communication in UAM. The unique challenges posed by aerial vehicles, such as signal interference from airframe structures and propeller rotations, require innovative solutions to optimize communication performance. Hu et al. \cite{hu2022communications} highlighted the impact of airframe structures and propeller rotations on communication links, emphasizing the importance of optimized airframe designs and antenna placements to mitigate these effects. Their findings also suggested that specific frequency bands can enhance transmission quality, offering practical guidelines for improving communication quality.

Pham et al. \cite{pham2023High} contributed to hardware innovation by introducing a leaky-wave antenna (LWA) array tailored for UAM. This design utilizes a spoof-surface-plasmon-polariton (SSPP) structure to achieve high gain, wide azimuth coverage, and low back-radiation in the 52.33-68.7 GHz frequency range. Its compact structure ensures 85\% ground clearance, enabling seamless integration with other electronic components. This antenna addresses the demanding operational requirements of UAM, supporting efficient and reliable communication in complex urban environments.

\vspace{-1mm}
\subsection{Enhancing Communication Capabilities}

Integrating radar and communication systems is a promising approach to improve situational awareness and operational safety in UAM. Jeong et al. \cite{jeong2024geometric} proposed a geometric sequence decomposition (GSD) approach for enhancing range and velocity estimation in orthogonal frequency-division multiplexing (OFDM) radar systems. By analyzing received data as geometric sequences, this method significantly improved target detection accuracy, even under challenging signal-to-noise ratios (SNRs).

Building on radar-based solutions, Schurwanz et al. \cite{schurwanz2023compressed} developed a compressive sensing (CS)-based obstacle detection and direction-of-arrival (DoA) estimation framework. The system uses a multiple-input multiple-output (MIMO) radar setup, achieving superior imaging performance compared to traditional methods. Validated against LiDAR data, the CS-based framework reduced data requirements while maintaining high accuracy, offering a robust tool for navigation in cluttered urban environments.

\section{UAM Management}
\label{Sec:management}

\subsection{Optimizing Traffic Flow and Scheduling}

\subsubsection{Traffic Demand Analysis}

Understanding traffic demand is critical for optimizing UAM operations and designing effective transportation solutions. Bulusu et al. \cite{bulusu2021traffic} analyzed commuter demand in the San Francisco Bay Area, focusing on UAM's potential to attract urban travelers by reducing travel time and optimizing vertiport transfer efficiency. The study revealed that during peak congestion, up to 45\% of commuters would prefer UAM, even with a long vertiport transfer time. Under less congested conditions, approximately 3\% of commuters could benefit from UAM services. The findings highlighted the strategic importance of vertiport placement and the need for UAM to address commuter preferences for time savings and accessibility. This research underscores the potential of UAM to transform urban transportation and emphasizes the importance of aligning UAM infrastructure with commuter behaviors.

\subsubsection{Scalable Scheduling Approaches}
Rigas et al. \cite{rigas2024scheduling} proposed an Integer Linear Programming (ILP) model for optimal scheduling, complemented by two heuristic algorithms to ensure scalability. The first algorithm incrementally solves the ILP for each vehicle, while the second employs a look-ahead search strategy. Both methods incorporate vehicle relocation to improve service coverage. Through realistic scenario evaluations, the study demonstrated the effectiveness of these approaches in balancing computational efficiency and optimality, paving the way for dynamic scheduling adaptations that account for operational constraints and commuter demands.

Expanding on collaborative management, Park et al. \cite{park2023multi} introduced a multi-agent deep reinforcement learning (MADRL) framework to enhance UAM traffic management. Using a centralized training and distributed execution (CTDE) approach, the algorithm enables cooperation among multiple vehicles, ensuring equitable service distribution without disadvantaging any individual vehicle. Simulations based on real-world vertiport maps validated the scalability and robustness of this framework. This work provided a feasible pathway for efficiently managing autonomous UAM operations in high-density urban environments.

\subsection{Ensuring Safety and Resilience in UAM Operations}

\subsubsection{Air Traffic Management Model}
To address the security challenge in densely populated urban airspace, Bharadwaj et al. \cite{bharadwaj2021decentralized} proposed a decentralized and hierarchical air traffic management (ATM) model to ensure safety and scalability. The framework organizes air traffic control through "vertihubs" and "vertiports", where the former manages local airspace and the latter oversees vehicle operations. A contract-based reactive synthesis approach ensures that each component operates independently while maintaining global safety standards. This decentralized structure not only supports high traffic volumes but also enhances system resilience.

\subsubsection{Dynamic Safety Mechanisms}
Wu et al. \cite{wu2022safety} addressed the challenge of collision avoidance in dense urban airspace by introducing a decentralized guidance algorithm. The algorithm models the problem as a multi-agent MDP with continuous action spaces, solved using an MCTS method. Enhanced by loss-of-chance constrained separation for safety assurance, the algorithm also employs Gaussian process regression and Bayesian optimization to discretize the action space, optimize flight time, and reduce mid-air collision risks. The proposed solution is capable of real-time operations and demonstrates significant scalability, marking a critical advancement in ensuring the safe operation of autonomous UAMs.

\subsubsection{Vertiport Safety Strategy}
The operational safety of the UAM is heavily dependent on the availability and reliability of vertiports. Wei et al. \cite{wei2023safe} proposed a method to verify the safety of UAM schedules, particularly in the circumstance of unexpected vertiport closures. By establishing necessary and sufficient conditions for schedule safety, the study introduced a linear programming-based algorithm that utilizes unimodular matrices to verify schedule feasibility efficiently. The algorithm accounts for uncertainties in travel time and ensures that rerouted vehicles do not exceed the capacity of alternative landing sites. Scalable to networks that involve up to 1,000 UAVs, this approach provides a robust tool to ensure operational safety and resilience, addressing the unpredictability of urban airspace management.

\section{Sustainability Implications in UAM}
\label{Sec:sustainability}

The rapid development of UAM necessitates addressing environmental issues that extend beyond traditional pollution control to encompass broader societal, environmental, and operational dimensions. 


Al Marzooqi et al. \cite{marzooqi2024electrical} presented a comprehensive review of eVTOL powertrain architectures, highlighting design trade-offs among wingless, lift plus cruise, and vectored thrust configurations for different mission profiles. The study discussed hybrid and fully electric powertrain by evaluating advanced energy storage technologies. Lithium-ion batteries, hydrogen fuel cells, and supercapacitors were investigated for reducing emissions and improving energy efficiency. Complementing this work, Alenezi et al. \cite{alenezi2025energy} proposed a hybrid energy storage system focusing on combining supercapacitors with lithium-ion batteries to enhance power management. Supercapacitors handle peak power demands during takeoff and landing, resulting in a 25.24\% reduction in total energy losses per flight while extending battery life. These improvements were validated through simulation and hardware-in-the-loop experiments, demonstrating their potential to increase energy efficiency and thereby contribute to the long-term environmental and economic sustainability of eVTOL operations.

Recognizing the variability of real-world urban operations, Hagag et al. \cite{hagag2025demand} developed a trajectory-based energy demand modeling framework incorporating non-nominal mission scenarios such as diversions, holding maneuvers, and vertiport closures. By integrating 3D Model Predictive Control (MPC)-based path planning with drivetrain-level simulations, the study quantified significant additional energy consumption, highlighting the need for adaptive energy reserve margins to ensure mission resilience and regulatory compliance in complex UAM environments. On the infrastructure side, Wang et al. \cite{wang2024sustain} explored how 5G-enabled smart city architectures support sustainable UAM integration by facilitating real-time coordination, energy management, and system-wide optimization across multimodal transportation networks. The study demonstrated the potential of 5G implementation in vehicle-to-grid (V2G) systems, electrified public transport, environmental monitoring, and traffic management. An integrated impact assessment framework was introduced to capture environmental, energy, transport, social, and economic sustainability metrics, emphasizing the cross-domain coordination required for sustainable urban mobility.

From a broader systems perspective, Zewde et al. \cite{zewde2025uam} investigated demand forecasting models for UAM deployment in the Atlanta Metro Region. Using multiple linear regression and random forest methods, the study identified population growth, income levels, congestion, and vertiport infrastructure as critical factors influencing long-term UAM demand scalability. The finding emphasized the need for coordinated economic growth, infrastructure development, and public affordability to ensure sustainable adoption of UAM services. Recent sustainability research has further expanded to explore human-centered aspects of autonomous transportation systems. In ground vehicle studies, researchers have examined interactions between autonomous vehicles and vulnerable road users from a sustainability perspective, focusing on safety, efficiency, and societal well-being \cite{he2024simulation,he2024risk,he2024sustainability, li2002automated, liu2023metamining, li2024sora, yin2023reliable, nazat2024xai, li2018automated}. Similar considerations are likely to extend to UAM systems as urban airspace becomes increasingly shared by diverse operators and stakeholders. 

Collectively, these studies demonstrate that sustainable UAM deployment requires holistic integration of energy-efficient propulsion systems, adaptive power management, resilient operational planning, scalable infrastructure, intelligent communication networks, and human-centered system design. Achieving long-term viability will therefore require coordinated advances across technical, infrastructural, and societal aspects.

\vspace{-2mm}
\section{Discussion}
\label{Sec:discussion}
\begin{figure*}[htbp]
\centerline{\includegraphics[width=1.0\linewidth]{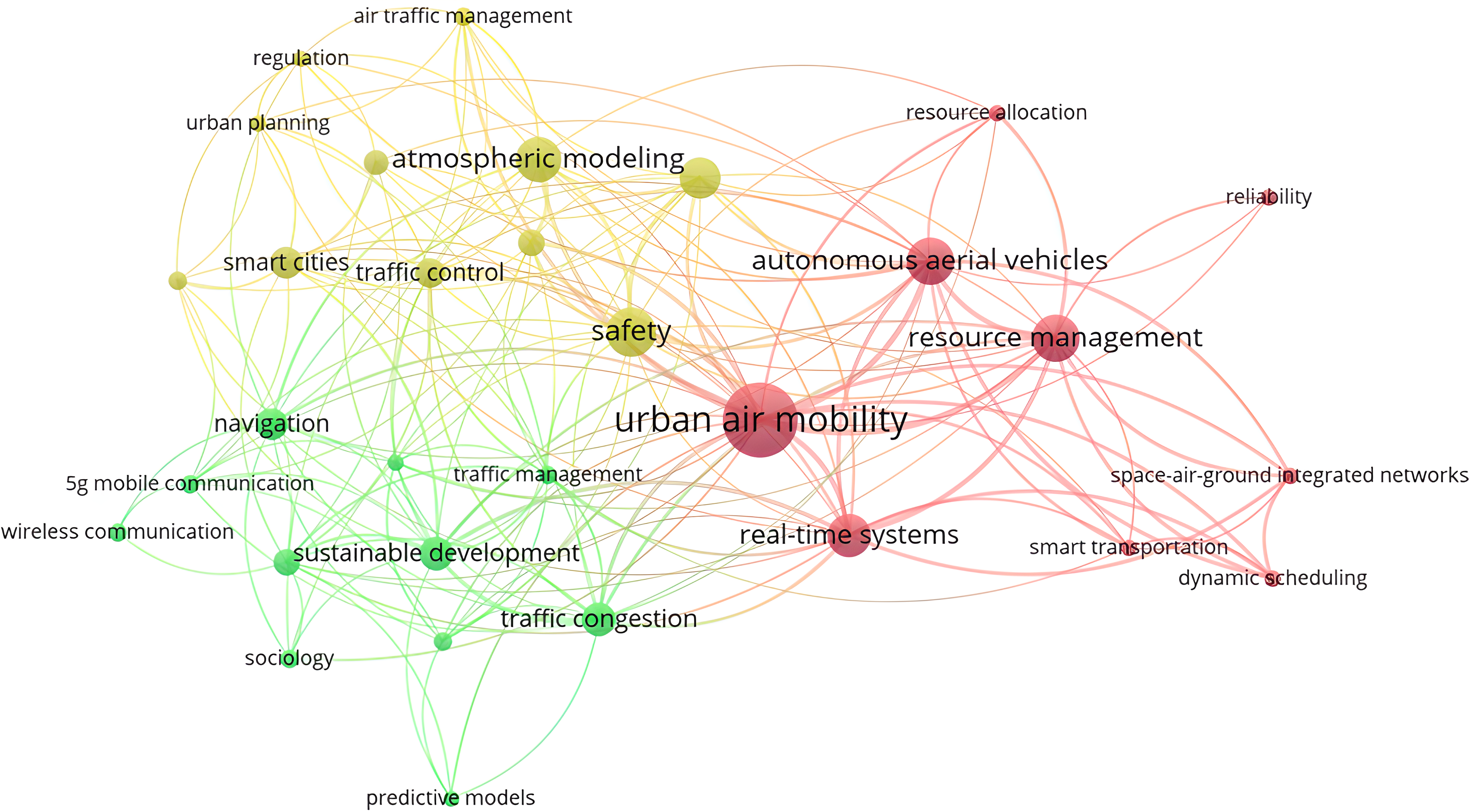}}
\vspace{-2mm}
\caption{Keywords map of post-2020 UAM research and development focusing on communication, management, and sustainability (generated by VOSviewer \cite{van2017citation}). Larger keywords indicate higher frequency of appearance in the reviewed literature.}
\label{fig:uam-keyword-analysis}
\end{figure*}

The integration of communication, management, and sustainability is essential for a safe, efficient, and scalable UAM ecosystem. Progress in any one domain directly shapes the feasibility and performance of the others. This section examines their interdependencies and implications for deployment.
\subsection{Interdependencies Between Communication, Management, and Sustainability}

Robust communication networks serve as the backbone for UAM management systems by enabling real-time coordination, adaptive scheduling, and data-driven decision-making. The cUAM framework \cite{han2022deep} supports reliable air–ground communication, while reinforcement learning-based spectrum management \cite{han2023joint} mitigates congestion and interference. These capabilities are prerequisites for resilient operations, including schedule safety verification under vertiport closures \cite{wei2023safe} and distributed turbulence prediction via federated learning \cite{zeng2023wireless}.

Sustainability goals further intensify these couplings. Management must incorporate dynamic energy reserves as diversions and holding patterns raise mission demand \cite{hagag2025demand}, and align vertiport allocation with projected demand and infrastructure growth \cite{zewde2025uam}. Advanced communications further contribute by coordinating charging, airspace use, and congestion mitigation within smart city platforms \cite{wang2024sustain}. At the hardware layer, energy-efficient propulsion \cite{marzooqi2024electrical} and hybrid energy storage \cite{alenezi2025energy} need tight integration with management policies and communications to maintain safety while meeting system-level sustainability targets.

\subsection{Toward an Integrated UAM Ecosystem}

The close interconnection of these domains emphasizes the necessity for a holistic system integration. Communication infrastructures not only enable real-time control but also serve as enablers for energy-aware traffic management and resilient scheduling strategies. Sustainability goals shape both infrastructure design and operational policies, driving the need for predictive management frameworks that account for fluctuating energy availability, demand patterns, and environmental conditions. The combination of these technologies is essential for creating a resilient UAM ecosystem capable of scaling to urban demand while minimizing environmental impact.

The keyword co-occurrence map generated from recent literature (shown in Fig.~\ref{fig:uam-keyword-analysis}), reinforces these insights. In the UAM field, \textit{safety}, \textit{autonomous aerial vehicles}, and \textit{resource management} are the most popular topics, as the main objective is to provide a resilient and efficient aerial transportation option. Following the top three, \textit{sustainable development} and \textit{atmospheric modeling} emerge as popular research focuses, while \textit{smart cities} and \textit{real-time systems} highlight the increasing integration between technological domains. The tight coupling of \textit{traffic management}, \textit{traffic congestion}, and \textit{navigation} reflects a growing consensus that achieving sustainable UAM requires multi-domain coordination and also depends on advanced communication and an optimized management architecture.

\vspace{-1mm}
\subsection{Future Expectations and Challenges}
Scalable deployment depends on tightly coupled advances across communication, management, and sustainability. Public acceptance will be shaped primarily by demonstrated safety, affordability, and equitable access. Meeting these expectations requires predictive, multi-agent management that maintains safety margins under variable demand and energy conditions. Reliable, low-latency communications are necessary conditions to support autonomous coordination and situational awareness. Energy-efficient propulsion with mission-aware power planning integrated into urban infrastructure is fundamental for a greener UAM operation.

Regulatory frameworks must evolve in parallel. Priorities include standardizing air-to-ground communication protocols and spectrum usage, establishing performance-based requirements for autonomy, and ensuring fleet-level safety. Additionally, efforts are needed to enhance cybersecurity and data privacy for networked operations, as well as to harmonize rules for vertiport siting and procedures. Interoperable architectures that align these regulatory, operational, and technical elements are pivotal to realizing a resilient and adaptive UAM ecosystem.

\section{Conclusion}
\label{Sec:conclusion}
\vspace{-2mm}
In this paper, we analyzed the key technological trends in UAM, focusing on communication, management, and sustainability. These interconnected areas were identified as critical enablers of secure, efficient, and expandable UAM systems. By examining recent advancements, we highlighted their contributions to the development of modern UAM operations and uncovered their interdependencies. This paper provided a foundation for understanding how these technologies can collectively shape the future of UAM, paving the way for further innovation in this transformative domain.

\bibliography{ref}

\end{document}